\nonstopmode
\documentclass[aip,apl,reprint,superscriptaddress]{revtex4-1}

 \usepackage{graphicx}
 \usepackage{color}
 \usepackage{epsfig}
 \usepackage{times}
 \usepackage{amsmath}
 \usepackage{amsthm}
 \usepackage{amsfonts}
 \usepackage{amssymb}
 \usepackage{braket}

\begin{document}

\title{Demonstration of vector magnetic field sensing by simultaneous control of nitrogen-vacancy centers in diamond using multi-frequency microwave pulses}

\author{Ken Yahata}
\affiliation{School of Fundamental Science and Technology, Keio University, 3-14-1 Hiyoshi, Kohoku-ku, Yokohama, Kanagawa 223-8522, Japan}
\author{Yuichiro Matsuzaki}
\affiliation{NTT Basic Research Laboratories, NTT Corporation, 3-1 Morinosato-Wakamiya, Atsugi, Kanagawa 243-0198, Japan}
\author{Shiro Saito}
\affiliation{NTT Basic Research Laboratories, NTT Corporation, 3-1 Morinosato-Wakamiya, Atsugi, Kanagawa 243-0198, Japan}
\author{Hideyuki Watanabe}
\affiliation{Correlated Electronics Group, Electronics and Photonics Research Institute, National
Institute of Advanced Industrial Science and Technology (AIST), Tsukuba Central 5, 1-1-1
Higashi, Tsukuba, Ibaraki 305-8565, Japan}
\author{Junko Ishi-Hayase}
\affiliation{School of Fundamental Science and Technology, Keio University, 3-14-1 Hiyoshi, Kohoku-ku, Yokohama, Kanagawa 223-8522, Japan}

\begin{abstract}
An ensemble of nitrogen-vacancy (NV) centers in diamond is a fascinating candidate to realize a sensitive
magnetic field sensor. In particular, since the axes of the NV centers are distributed along four directions, a collection
of  measurement data from  NV centers with different axes provides information on the vector
components of a magnetic field. However, in the conventional approach, the low measurement contrast of
 NV centers limits the sensitivity of vector magnetic field sensing. Recently, to overcome this problem, multi-frequency control of the NV centers has been proposed. The key idea is that the four types of
NV centers with different axes are simultaneously controlled by multi-frequency microwave pulses. Here, we
demonstrate vector magnetic field sensing with an ensemble of NV centers in diamond via
such  multi-frequency control with pulsed-type measurements. We use Hahn echo pulses and  extract  information on the vector components of an applied AC magnetic field. We
find that the sensitivity of  diamond-based vector field sensing
 with  multi-frequency control is
 better than that with single-frequency control for every vector
 component of a magnetic field.
\end{abstract}

\maketitle


Magnetic field sensors have significant applications in 
chemistry, biology, and medical science. For example, in electron spin
resonance, which is a widely used technique in chemistry, 
magnetic field sensors play an important role in obtaining  information about
the electron spin. Magnetoencephalography is a clinical technique for measuring
 electrical activity  in the brain via magnetic field sensing, thereby providing information about 
brain function. Magnetic
resonance imaging allows examination of the internal structure of the
human body based on magnetic field information. For the magnetic field sensors used in these contexts, sensitivity
and spatial resolution are  essential parameters to quantify performance, and much effort
has been devoted to measuring weak magnetic fields in local
regions.\cite{Simon,Chang,Poggio}

Qubit-based sensors are attractive candidates for use in such applications. A solid-state qubit can
be coupled with a magnetic field, resulting in a shift in the resonant frequency
of the qubits. The associated energy shift can be detected via Ramsey measurements in the case of an 
applied DC magnetic field. It is also possible to detect an AC magnetic field by using Hahn echo pulse sequences,
which suppress dephasing of the qubits,
thereby typically improving sensitivity. A number of types of qubit sensor have been demonstrated experimentally, using, for example, atoms,
superconducting flux qubits, or ion traps.

One important type of qubit-based sensor uses nitrogen-vacancy (NV) centers in diamond.\cite{Maze,Taylor,Balasu} The NV center is a spin-1 system, and
magnetic fields can change the energies of the states  $\ket{\pm{1}}$. By
using frequency selectivity, we can consider an NV center
as an effective two-level system comprising $\ket{0}$ and $\ket{1}$, with
$\ket{-1}$ being energetically well detuned from the other states. Since
 NV centers can be strongly coupled with optical photons,\cite{Gruber,Jelezko}
it is possible to initialize and read out the spin state of $\ket{0}$ and
$\ket{1} (\ket{-1})$ by applying a green-light laser. High-fidelity gate operations
with NV centers have been demonstrated using
microwave pulses.\cite{Davies,Gruber,Jelezko,Jelezko2} Long coherence times in the region of a few milliseconds have been observed even at room temperature.\cite{Balasu2,Mizuochi,Bar} Owing to
these properties,  NV centers can be considered as suitable
systems to use in sensitive magnetic field sensors.

\begin{figure}[t]
\begin{center}
  \includegraphics[width=6.5cm]{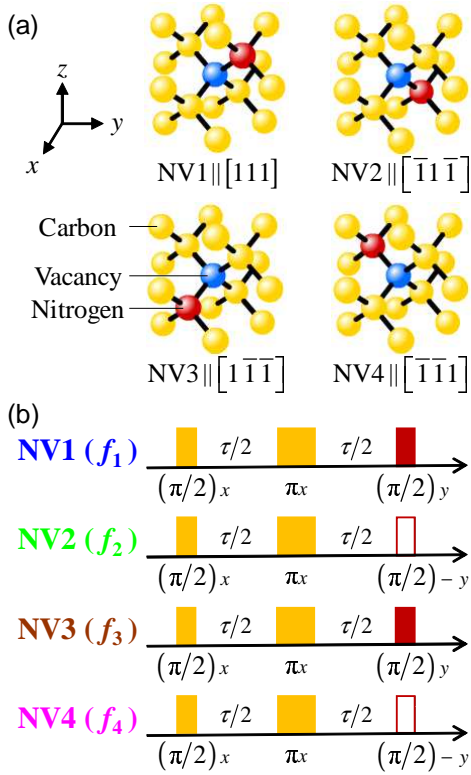} 
  \caption{
(a) The structure of an NV center in diamond. This contains a
 nitrogen atom (substituting for a carbon atom) and a vacancy in the
 diamond lattice sites. NV centers have four possible crystallographic axes.
(b) Microwave pulse sequence for multi-frequency control of NV centers.
 In this case, the phase of the last microwave pulse is shifted by
 $180^{\circ}$ for  NV2 and NV4 to
 extract  information on the $x$ component of the vector
 magnetic field.
Similarly, to extract the $y$ ($z$) component, the same phase shift of the 
 microwave pulse is required for  NV3 and NV4 (NV2 and NV3).
\label{fig1}
 }
\end{center}
\end{figure}

Moreover, NV centers are useful for the measurement of the vector components of applied
magnetic fields.\cite{Maertz,Steinert,Pham,Tetienne,Dmitriev,Sasaki} As can be seen in Fig.~\ref{fig1} (a), NV centers have four possible crystallographic axes:
$[111]$ (NV1), $[\bar{1}1\bar{1}]$ (NV2), $[1\bar{1}\bar{1}]$ (NV3), and $[\bar{1}\bar{1}1]$ (NV4).
The vector component of a magnetic field along the direction of one of these axes will induce a frequency shift in an NV center with that axis. Thus, a collection of the measurement data from the four
types of NV center with different axes provides information on  all the vector components of a magnetic field.
There have been a number of experimental demonstrations of the use of NV
centers to measure  vector components of magnetic fields. For example, 
vector components of AC magnetic fields from current-carrying wires have been measured and analyzed using this technique,\cite{Pham} and vector 
imaging of DC magnetic fields
from living cells and circuit currents has been demonstrated.\cite{Sage,Nowodzinski}
Attempts have also been made to increase
the signal-to-noise ratio of vector measurement
by combined RF excitation of hyperfine triplets in the optically detected magnetic resonance (ODMR) spectrum.\cite{Dmitriev}

However, such diamond-based sensors are affected by low
readout contrast, which significantly decreases their sensitivity for vector fields.\cite{Taylor,Acosta} Although photoluminescence
(PL) from the state  $\ket{0}$ is slightly larger than that from the
state  $\ket{1}$, most of the photons emitted from the NV centers are absorbed in the environment, and so  
a large number of repetitions are typically needed to determine the population of the state $\ket{0}$ or $\ket{1}$ from the optical readout. Moreover, in the
conventional approach to  vector field sensing, during measurement of any one of the four types of  NV center, the other three types
remain in the state  $\ket{0}$ regardless of the applied
magnetic field, which induces additional noise.

Recently, to overcome this problem of  low readout
contrast, an approach to vector magnetic field sensing 
has been proposed in which the four types
of NV center with different axes are simultaneously controlled by multi-frequency microwave pulses.\cite{Kitazawa,Walsworth,Zhang} Consequently, all the NV centers irradiated by a green laser  contribute to
the signal  containing  information on the magnetic field, in contrast to the conventional approach, in which $75\%$ of the NV centers just
induce noise without contributing to the signal. 
The sensitivity of this method of vector field sensing
can, in
principle, be four times higher than that of the conventional approach.
It is worth mentioning that vector magnetic field
sensing with simultaneous control of  NV centers has been
achieved  using CW-ODMR measurements,\cite{Walsworth,Zhang} 
but multi-frequency control using pulsed-type measurements (which
typically have the advantage that they can suppress dephasing) has not yet been demonstrated.

In this paper, we demonstrate  vector magnetic
field sensing with multi-frequency control.
We adopt a spin echo measurement technique, which is one of the typical pulsed ODMR
schemes for measuring AC magnetic fields using NV centers.
In a spin echo measurement, by irradiating NV centers with the
microwave pulse sequence
shown in Fig.~\ref{fig1} (b),
the AC magnetic field induces a
phase shift on the state of
the electron spin. By reading out the state using the green laser, we can
detect the amplitude of the AC magnetic field. In  multi-frequency
control vector magnetic
field sensing, we simultaneously measure signals from the four NV axes by using
microwave pulses with four different frequencies. It is important for increased sensitivity that  the signals be
naturally synthesized when we read out the state  optically. For this purpose, by applying an
external static
magnetic field, we separate the resonant frequencies of the four axes so
that we can individually control the states of the NV centers with
different axes.
If the  microwave
pulses were all the same, the optical signals from  NV centers with different axes would cancel,  decreasing
the sensitivity. The key idea in our scheme is to shift the phase of the
last microwave pulse by $180^{\circ}$ for  NV centers with specific
axes by using frequency
selectivity. This operation will reverse the
spin direction, depending on the direction of the axis, and the
optical signal from the 
NV centers with that axis will be enhanced. Figure~\ref{fig1} (b) shows the appropriate sequence for the $x$ component, and, as indicated in the caption, a similar procedure can be adopted to measure the $y$ and $z$ components. The details of the theoretical analysis are described elsewhere\cite{Kitazawa}.

\begin{figure}[t]
\begin{center}
  \includegraphics[width=6.5cm]{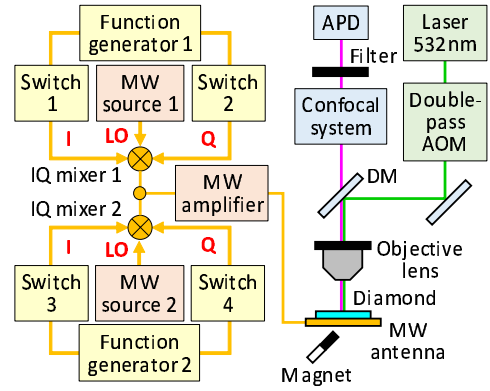} 
  \caption{Experimental setup.
 Microwaves (MW) are generated by two sources, and are modulated by plus
 and minus  sidebands at radiofrequency from a function generator 
 to produce multiple frequencies for the control of  NV
 centers with  four different axes.
 They are amplified,
 combined, and radiated by a MW antenna.\cite{Sasaki2}
 MW pulses are generated by four switches, which are controlled by a pulse
 blaster. The excitation laser beam passes through an acousto-optic modulator
 (AOM), which is used to switch the laser. After the AOM, the beam passes through
 an optical fiber and objective lens, whose position is controlled by a
 piezo stage.
 PL from the NV center passes through a confocal system and long-pass
 filter at 630\,nm and is collected by an avalanche photodiode (APD).
 The target AC magnetic field is generated by a copper wire.\label{fig2}}
\end{center}
\end{figure}

Figure~\ref{fig2} shows our experimental setup for multi-frequency control vector
magnetic field sensing. We use a home-built system for
confocal laser scanning
microscopy with a spatial resolution of 400\,nm. The diamond sample is
positioned above the antenna used to generate the microwaves. A
magnet is placed below the
antenna\cite{Sasaki2} to apply a static magnetic field, and a copper wire is placed in
contact with the sample surface to apply the target AC magnetic
field. We use a single-photon-resolving detector to measure the photons from the NV centers. Microwaves of four
different frequencies  are generated by modulating the microwaves from
two  sources,  each of which gives
two frequencies, using in-phase/quadrature (IQ) mixers and function generators. The microwave phase ``x'' is controlled
by switches 1 and 3 (LO + I) and the  phase ``y''  by switches 2 and 4 (LO + Q).

\begin{figure}[t]
\begin{center}
  \includegraphics[width=6.5cm]{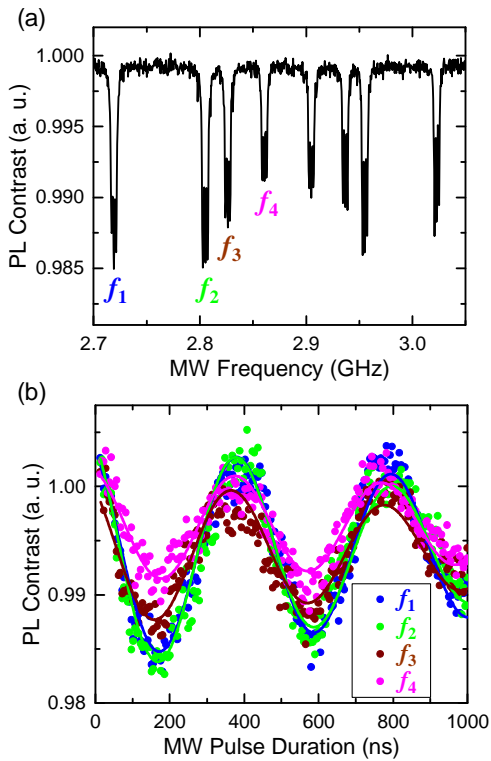} 
  \caption{(a) Experimental ODMR spectrum. In an applied
 magnetic field, we observe eight resonances,  corresponding to the
 four types of  NV centers with different axes. This
 frequency selectivity can be used for individual control of the different
 NV centers. (b) Rabi oscillations of the NV centers with four different axes.
The distribution
 ratios of the  different NV centers are determined from these
 Rabi oscillations.\label{fig3}}
\end{center}
\end{figure}

The details of the diamond sample used in our
experiment are as follows. We use an ensemble of NV centers in a $2\,\mu$m-thick diamond
film on a (001) electronic-grade
substrate. Diamond films were prepared from an isotopically-enriched ($>$ 99.999\% for $^{12}$C) $^{12}$CH$_{4}$, H$_{2}$ mixed gas system in a microwave plasma-assisted chemical vapor deposition.
To both increase the NV center density and improve the coherence time,
the sample was irradiated with 350\,keV He$^+$ ions at ion doses of
$10^{12}$\,cm$^{-2}$,
followed by annealing for 24\,hours in vacuum at $800\,^\circ$C.\cite{Heion} To ensure a
uniformly random distribution about the four possible crystallographic
axes, with each
orientation rate being around 25\%, the sample was annealed for 15\,hours in vacuum at $1200\,^\circ$C.\cite{reorientation,reorientation2} The NV center density of this diamond sample is
$2.2\times10^{15}$\,cm$^{-3}$ and the coherence time is 20\,$\mu$s.

\begin{figure}[t]
\begin{center}
  \includegraphics[width=6.5cm]{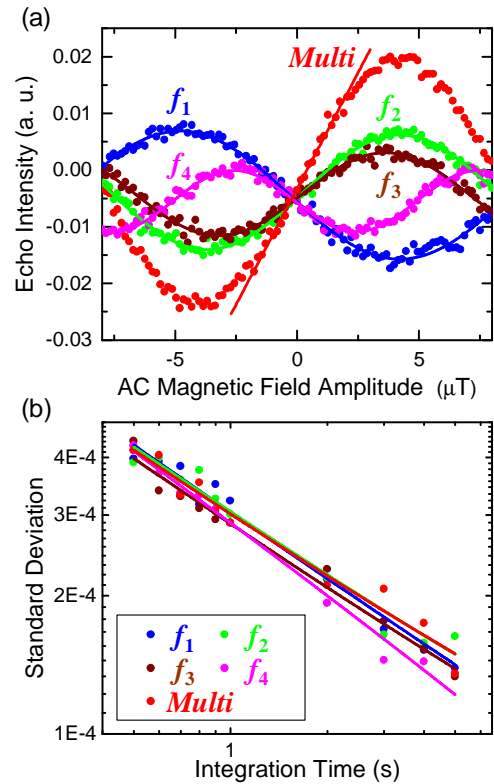} 
  \caption{(a)
 Spin echo signals vs. applied AC magnetic field amplitude using standard
 single-frequency control and our multi-frequency control method. Multi-frequency
 control clearly leads to enhancement of the signal. From these
 behaviors, we can determine the signal gradient
 $dP_{n}/dB_{n}$ against the amplitude of the AC magnetic field.
Here, to enhance the optical signal, the phase of the readout microwave pulse is shifted by $180^{\circ}$ for $f_1$ and $f_4$.
 (b) Standard deviation $\delta P_{n}$ of the signal. As we increase
 the integration time $T$, the standard deviation decreases as 
 $\delta P\propto 1/\sqrt{T}$ up to around $T\simeq 1$\,s.\label{fig4}}
\end{center}
\end{figure}

We demonstrate  vector AC magnetic field sensing with a frequency of
100\,kHz by  multi-frequency control.
First, we measure the ODMR
spectrum and Rabi
oscillations, as shown in Fig.~\ref{fig3}. An applied static magnetic field 
separates the frequencies of  NV centers with the four different axes, and we
adjust the Rabi
frequency around 2.5\,MHz (which corresponds to a time scale for the $\pi $
pulse of around 200\,ns). From the ODMR spectrum, we determine the
resonant frequencies of
the NV centers with different axes: $f_1=2.72$\,GHz (NV1 $= [111]$),
$f_2=2.806$\,GHz (NV2 $= [\bar{1}1\bar{1}]$), $f_3=2.826$\,GHz
(NV3 $=[1\bar{1}\bar{1}]$), and $f_4=2.862$\,GHz (NV4 $=[\bar{1}\bar{1}1]$). From the behavior of the Rabi oscillations, we determine the orientation ratios as (NV1)~:~(NV2)~:~(NV3)~:~(NV4) = 29\%~:~35\%~:~21\%~:~15\%.

Second, we consider a particular case and measure the amplitude of
an AC magnetic field under the assumption that its direction  is known.
To find the sensitivity of the conventional scheme,
we measure
the optical signal against the magnetic field amplitude and also
measure the signal fluctuation
against the repetition time when only a single-frequency microwave is used
to control the NV centers. 
These results are shown in Fig.~\ref{fig4} ($f_1,\ldots,
f_4$). The uncertainty of the estimation is
given by $\delta
B_{n}=\delta P_{n}/(dP_{n}/dB_{n})$ ($n=f_1, \ldots,
f_4, \rm{multi}$), 
and we
obtain the sensitivity of each axis from Fig.~\ref{fig4} as follows (units of nT/$\sqrt{\rm{Hz}}$):  $80\pm11$
(NV1), $76\pm8$ (NV2), $85\pm11$(NV3), and $85\pm8$ (NV4). We then perform similar measurements to determine the sensitivity of our
approach with multi-frequency control of the NV centers.
Here, to enhance the optical signal, we shift the phase of the last
microwave pulse by $180^{\circ}$ for $f_1$ and $f_4$ (NV1 and NV4), so that
the sign of the signals
from NV1 and NV4 around $0\,\mu$T becomes the same as the sign of those from NV2 and
NV4. These results are also shown in Fig.~\ref{fig4}. We obtain the sensitivity
of measurement of the AC magnetic
field with our scheme
as $\delta B_{\rm{mf}}=35\pm5\,\rm{nT}/\sqrt{\rm{Hz}}$, which is
approximately twice as good as the sensitivity of the conventional scheme.

\begin{figure}[t]
\begin{center}
  \includegraphics[width=6.5cm]{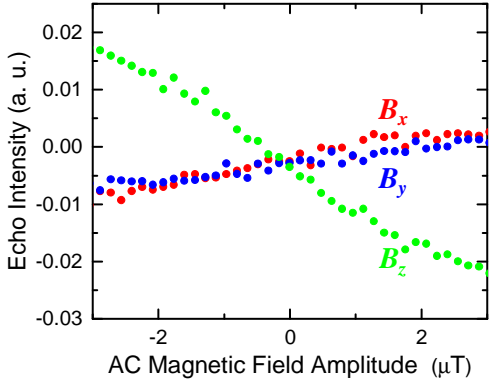} 
  \caption{Demonstration of multi-frequency control vector magnetic field sensing. In this study, to measure $B_{x}$, $B_{y}$, and $B_{z}$, we shift the phase of the last microwave pulse by $180^{\circ}$ for $f_2$ and $f_4$ (NV2 and NV4), $f_3$ and $f4$ (NV3 and NV4), and $f_2$ and $f_3$ (NV2 and NV3).\label{fig5}}
\end{center}
\end{figure}

Finally, we measure the vector components of the AC magnetic field under the assumption that we do not know
the direction of the field. In  Fig.~\ref{fig5}, we plot the echo intensity against the
amplitude of the AC magnetic
field for three different pulse sequences  corresponding to the
extraction of the three vector components $B_{x}$, $B_{y}$, and $B_{z}$, respectively. Using our multi-frequency control scheme, we obtain
$(\tilde{B}_{x}, \tilde{B}_{y}, \tilde{B}_{z})=(0.23, 0.16, -0.97)$,
where the tilde $\tilde{ \ }$ denotes the dimensionless normalized vector component.
For  comparison, we also measure the vector components
using conventional single-frequency control and obtain nearly the same result:
$(\tilde{B}_{x}, \tilde{B}_{y}, \tilde{B}_{z})=(0.23, 0.17, -0.96)$. This demonstrates the ability of our
multi-frequency control method
to measure each vector component of an applied magnetic fields. Also,
we compare the sensitivities of measurement of $B_{x}, B_{y}$, and $B_{z}$ between
our scheme and
the conventional one. We define the uncertainties in the estimation of
the vector components using the
conventional scheme and our multi-frequency scheme as 
\begin{align}
\delta B^{\rm{(c)}}_{x}&=\frac {\sqrt{2}\,\delta P_{\rm{single}}}{( dP_{[111]}
 -dP_{[1\bar{1}\bar{1}]})/dB_{x}},
 \label{conv}
 \\
\delta B^{\rm{(mf)}}_{x}&=\frac {\delta P_{\rm{multi}}}{dP_{B_{x}}/dB_{x}},  \label{ourmulti}
\end{align}
where $P$ denotes the optical signals from the NV centers.
We  define the sensitivities for the $y$ and
$z$ components similarly.
It is worth mentioning that
to measure one of the vector components of the magnetic field using the
conventional method,
measurements of two signals from NV1 ($[111]$) and
NV3 ($[1\bar{1}\bar{1}]$) are required, which doubles the measurement
time. This is the reason for the factor 
$\sqrt{2}$ in the numerator of the
sensitivity for the conventional scheme in  Eq.~\eqref{conv}.
From Fig.~\ref{fig4} ($f_1,\ldots,f_4$), in the conventional scheme, the vector sensitivities
are obtained as as follows (units of nT/$\sqrt{\rm{Hz}}$):
$\delta B^{\rm{(c)}}_{x}=120\pm30$, $\delta
B^{\rm{(c)}}_{y}=200\pm60$, and 
$\delta B^{\rm{(c)}}_{z}=41\pm6$. In our
scheme, the vector sensitivities are 
 obtained from Fig.~\ref{fig5} as follows (units of nT/$\sqrt{\rm{Hz}}$): $\delta B^{\rm{(mf)}}_{x}=35\pm10$, 
$ \delta B^{\rm{(mf)}}_{y}=42\pm16$, and $ \delta B^{\rm{(mf)}}_{z}=34\pm5$.
 Therefore, we conclude that our scheme shows better sensitivity than the conventional scheme for every vector component of the AC magnetic field.
 In these demonstrations, $\delta B^{\rm{(mf)}}_z$ is
 close to $\delta B_{\rm{mf}}$. When we measure $\delta B_{\rm{mf}}$,
 the magnetic field direction is near to the $z$ direction, and so these
 experimental results are consistent in our setup.

The theoretical ratios between the sensitivities of our scheme and those of the
conventional scheme for each vector magnetic field component can be
calculated  as follows:\cite{Kitazawa} $\delta B_{x}^{\rm{(c)}}/\delta
 B_{x}^{\rm{(mf)}}=\delta B_{y}^{\rm{(c)}}/\delta
 B_{y}^{\rm{(mf)}}=\delta B_{z}^{\rm{(c)}}/\delta
 B_{z}^{\rm{(mf)}}\simeq 4$,
and thus the ratio should be
the same for each magnetic field component.
However, in our
experiment,
the sensitivity improvement ratio depends on which vector component we have extracted for the
vector field sensing.
 One possible reason is imperfect application of
the microwave pulses with four frequencies in our setup. The use of
a microwave antenna  to recover the sensitivity could
solve this problem.
The discrepancy could also be due to the fact that the ratios of the crystallographic
orientations are not equal among the four directions in our experiment, whereas the theoretical
calculations assume a 25\% orientation ratio for all four
directions.
Such an inhomogeneous distribution of  orientations means that
the sensitivity depends on which  vector component is measured in our scheme.
In principle, we could overcome this problem by more sophisticated
sample fabrication. Moreover, adjusting the pulse
sequence period
could, in principle, change the effective orientation ratio.\cite{Kitazawa}

In conclusion, we have demonstrated  vector magnetic field sensing with
multi-frequency control of  NV centers in diamond. We have achieved a
sensitivity twice
better than that of the conventional scheme. By choosing an
appropriate pulse sequence, we can extract the vector components of a
magnetic field.
Although we use a spin echo method in our demonstration in this paper, the
technique described here is quite general and  could be applied with
other methods, such as
DC magnetic field sensing with Ramsey measurement or AC magnetic field sensing with dynamical decoupling for further improvement in sensitivity. 

This work was supported by JSPS KAKENHI (Grant No. 15K17732), MEXT KAKENHI (Grants Nos. 18H01502, 15H05868, 15H05870, 15H03996, 26220602, and 26249108), SENTAN.JST, and Spin-NRJ.


\begin{thebibliography}{30}
\bibitem{Simon}J. Simon, Adv. Phys. \bf48\rm, 449 (1999).
\bibitem{Chang}A. Chang, H. Hallen, L. Harriott, H. Hess, H. Kao, J. Kwo,
R. Miller, R. Wolfe, J. Van der Ziel, and T. Chang, Appl. Phys. Lett. \bf61\rm, 1974 (1992).
\bibitem{Poggio}M. Poggio, and C. Degen, Nanotechnol. \bf21\rm, 342001 (2010).
\bibitem{Maze}J. Maze, P. Stanwix, J. Hodges, S. Hong, J. Taylor, P. Cappellaro,
L. Jiang, M. Dutt, E. Togan, A. Zibrov, A. Yacoby, R. L. Walthworth, and M. D. Lukin, Nature \bf455\rm, 644 (2008).
\bibitem{Taylor}J. Taylor, P. Cappellaro, L. Childress, L. Jiang, D. Budker,
P. Hemmer, A. Yacoby, R. Walsworth, and M. Lukin, Nat. Phys. \bf4\rm, 810 (2008).
\bibitem{Balasu}G. Balasubramanian, I. Chan, R. Kolesov, M. Al-Hmoud,
J. Tisler, C. Shin, C. Kim, A. Wojcik, P. Hemmer, A. Krueger, T. Hanke, A. L. Storfer, R. Bratschitsch, F. Jelezko, and J. Wrachtrup, Nature \bf455\rm, 648 (2008).
\bibitem{Schaffry}M. Schaffry, E. Gauger, J. Morton, and S. Benjamin, Phys. Rev.
Lett. \bf107\rm, 207210 (2011).
\bibitem{Davies}G. Davies, ed., \it Properties and Growth of Diamond,  \rm (INSPEC, The institution of Electrical Engineers, London, 1994).
\bibitem{Gruber}A. Gruber, A. Dr{\"a}benstedt, C. Tietz, L. Fleury, J. Wrachtrup,
and C. V. Borczyskowski, Science \bf276\rm, 2012 (1997).
\bibitem{Jelezko}F. Jelezko, I. Popa, A. Gruber, C. Tietz, J. Wrachtrup, A. Nizovtsev,
and S. Kilin, Appl. Phys. Lett. \bf81\rm, 2160 (2002).
\bibitem{Jelezko2}F. Jelezko, T. Gaebel, I. Popa, A. Gruber, and J. Wrachtrup,
Phys. Rev. Lett. \bf92\rm, 076401 (2004).
\bibitem{Balasu2}G. Balasubramanian, P. Neumann, D. Twitchen, M. Markham,
R. Kolesov, N. Mizuochi, J. Isoya, J. Achard, J. Beck, J. Tissler, V. Jacques, P. R. Hemmer, F. Jelezko, and J. Wrachtrup, Nat. Mater. \bf8\rm, 383 (2009).
\bibitem{Mizuochi}N. Mizuochi, P. Neumann, F. Rempp, J. Beck, V. Jacques, P. Siyushev, K. Nakamura, D. Twitchen, H. Watanabe, S. Yamasaki, F. Jelezko, and J. Wrachtrup, Phys. Rev. \bf B 80\rm, 041201 (2009).
\bibitem{Bar}N. Bar-Gill, L. M. Pham, A. Jarmola, D. Budker, and R. L. Walsworth, Nat. Commun. \bf4\rm, 1743 (2013).
\bibitem{Maertz}B. Maertz, A. Wijnheijmer, G. Fuchs, M. Nowakowski, and
D. Awschalom, Appl. Phys. Lett. \bf96\rm, 092504 (2010).
\bibitem{Steinert}S. Steinert, F. Dolde, P. Neumann, A. Aird, B. Naydenov,
G. Balasubramanian, F. Jelezko, and J. Wrachtrup, Rev. Sci. Instrum. \bf81\rm, 043705 (2010).
\bibitem{Pham}L. M. Pham, D. Le Sage, P. L. Stanwix, T. K. Yeung, D. Glenn,
A. Trifonov, P. Cappellaro, P. Hemmer, M. D. Lukin, and H. Park, New J. Phys. \bf13\rm, 045021 (2011).
\bibitem{Tetienne}J. Tetienne, L. Rondin, P. Spinicelli, M. Chipaux, T. Debuisschert,
J. Roch, and V. Jacques, New J. Phys. \bf14\rm, 103033 (2012).
\bibitem{Dmitriev}A. K. Dmitriev, and A. K. Vershovskii, J. Opt. Soc. Am. B \bf33\rm, B1 (2016).
\bibitem{Sasaki}K. Sasaki, Y. Monnai, S. Saijo, R. Fujita, H. Watanabe, J. Ishi-Hayase, K. M. Itoh, and E. Abe, Rev. Sci. Instrum. \bf87\rm, 053904 (2016).
\bibitem{Sage}D. Le Sage, K. Arai, D. Glenn, S. DeVience, L. Pham, L. Rahn-Lee, M. Lukin, A. Yacoby, A. Komeili, and R. Walsworth, Nature \bf496\rm, 486 (2013).
\bibitem{Nowodzinski}A. Nowodzinski, M. Chipaux, L. Toraille, V. Jacques, J.-F. Roch, and T. Debuisschert, Microel. Rel. \bf55\rm, 1549 (2015).
\bibitem{Acosta}V. Acosta, E. Bauch, M. Ledbetter, C. Santori, K.-M. Fu, P. Barclay,
R. G. Beausoleil, H. Linget, J. F. Roch, F. Treussart, S. Chemerisov, W. Gawlik, and D. Budker, Phys. Rev. \bf B 80\rm, 115202 (2009).
\bibitem{Kitazawa}S. Kitazawa, Y. Matsuzaki, S. Saijo, K. Kakuyanagi, S. Saito, and J. Ishi-Hayase, Phys. Rev. \bf A 96\rm, 042115 (2017).
\bibitem{Walsworth}J. M. Schloss, J. F. Barry, M. J. Turner, and R. L. Walsworth, Phys. Rev. Appl. \bf10\rm, 034044 (2018).
\bibitem{Zhang}C. Zhang, H. Yuan, N. Zhang, L. Xu, J. Zhang, B. Li, and J. Fang, J. Phys. D: Appl. Phys. \bf51\rm, 155102 (2018).
\bibitem{Sasaki2}K. Sasaki, E. E. Kleinsasser, Z. Zhu, W. D. Li, H. Watanabe, K. M. C. Fu, K. M. Itoh, and
E. Abe, Appl. Phys. Lett. \bf110\rm, 192407 (2017).
\bibitem{Heion}E. E. Kleinsasser, M. M. Stanfield, J. K. Q. Banks, Z. Zhu,W.-D. Li, V. M. Acosta, H.Watanabe,
K. M. Itoh, and K.-M. C. Fu, Appl. Phys. Lett. \bf108\rm, 202401 (2016).
\bibitem{reorientation}H. Pinto, R. Jones, D. W. Palmer, J. P. Gross, P. R. Briddon, and S. {\"O}berg, Phys. Status Solidi A \bf209\rm, 1765 (2012).
\bibitem{reorientation2}T. Karin, S. Dunham, and K. M. Fu, Appl. Phys. Lett. \bf105\rm, 053106 (2014).
\end{thebibliography}
\end{document}